# The potassium absorption on HD189733b and HD209458b

Engin Keles,[1] Matthias Mallonn,[1] Carolina von Essen,[2] Thorsten A. Carroll,[1] Xanthippi Alexoudi,[1] Lorenzo Pino,[3] Ilya Ilyin,[1] Katja Poppenhäger,[1] Daniel Kitzmann,[4] Valerio Nascimbeni,[5,6] Jake D. Turner,[7,8] Klaus G. Strassmeier[1]

[1] *Leibniz-Institut für Astrophysik Potsdam (AIP), An der Sternwarte 16, 14482 Potsdam, Germany*
[2] *Stellar Astrophysics Centre, Department of Physics and Astronomy, Aarhus University, Ny Munkegade 120, 8000 Aarhus C, Denmark*
[3] *Leiden Observatory, Leiden University, Postbus 9513, 2300 RA Leiden, The Netherlands*
[4] *University of Bern, Center for Space and Habitability, Gesellschaftsstrasse 6, CH-3012, Bern, Switzerland*
[5] *Istituto Nazionale di Astrofisica, Osservatorio Astronomico di Padova, 35122 Padova, Italy*
[6] *Dipartimento di Fisica e Astronomia - Universta di Padova, Vicolo dell Osservatorio 3, I-35122 Padova*
[7] *Cornell University, Ithaca, New York, USA;* [8] *University of Virginia, Charlottesville, Virginia, USA*

**ABSTRACT**

In this work, we investigate the potassium excess absorption around 7699Å of the exoplanets HD189733b and HD209458b. For this purpose, we used high spectral resolution transit observations acquired with the 2 × 8.4m Large Binocular Telescope (LBT) and the Potsdam Echelle Polarimetric and Spectroscopic Instrument (PEPSI). For a bandwidth of 0.8Å, we present a detection $> 7\text{-}\sigma$ with an absorption level of 0.18 % for HD189733b. Applying the same analysis to HD209458b, we can set 3-$\sigma$ upper limit of 0.09%, even though we do not detect a K- excess absorption. The investigation suggests that the K- feature is less present in the atmosphere of HD209458b than in the one of HD189733b. This comparison confirms previous claims that the atmospheres of these two planets must have fundamentally different properties.

## 1 INTRODUCTION

A suitable method for the characterization of planetary atmospheres is transmission spectroscopy (Seager & Sasselov 2000). During transit, a small fraction of the starlight is absorbed or scattered by atoms and molecules, putting a fingerprint on the stellar spectrum. One possibility to infer these fingerprints is the "excess absorption" method, where the flux of the spectral range of interest is integrated within a bandwidth and divided by the flux within a reference band (a spectral region where planetary absorption is not expected), showing the absorption within the atmosphere during the transit (Charbonneau et al. 2002). Especially hot giant planets are suitable targets of these kind of investigations due to their large scale heights and short orbital periods, where several atmospheric constituents have been successfully detected e.g. sodium (Redfield et al. 2008; Snellen et al. 2008; Casasayas-Barris et al. 2017), potassium (Sing et al. 2011), titanium and iron (Hoeijmakers et al. 2018), hydrogen (Yan & Henning 2018), and magnesium (Cauley et al. 2019).
For hot Jupiter atmospheres with T ∼ 1500K, the strongest atomic absorber in the optical wavelengths are Na and K (Fortney et al. 2010). Different investigations on HD189733b and HD209458b have confirmed the presence of Na using low- and high resolution spectroscopy (see e.g. for HD209458b Charbonneau et al. (2002) and Snellen et al. (2008) or for HD189733b Redfield et al. (2008) and Wyttenbach et al. (2015)). However, the detection of K was not yet assured from high resolution investigations for any exoplanet, although attempts were made e.g. recently by Gibson et al. (2019) investigating K on the exoplanet Wasp-31b. Several investigations attempted to detect K on HD189733b and HD209458b. For instance, Jensen et al. (2011) used the Hobby-Eberly-Telescope and stated a non-detection of K for both exoplanets. A tentative 2.5-$\sigma$ detection of K in the atmosphere of HD189733b was claimed by Pont et al. (2013) using the ACS camera at the Hubble Space Telescope. To date, there is no significant detection of K for these two exoplanets, neither in low- nor in high- resolution observations.

## 2 OBSERVATIONS

We observed one transit for HD189733b on October 11, 2017 at 01:47 – 06:39 UT (PI: J.D. Turner, UVA) and one transit for HD209458b on October 13, 2017 at 03:03 – 08:04 UT

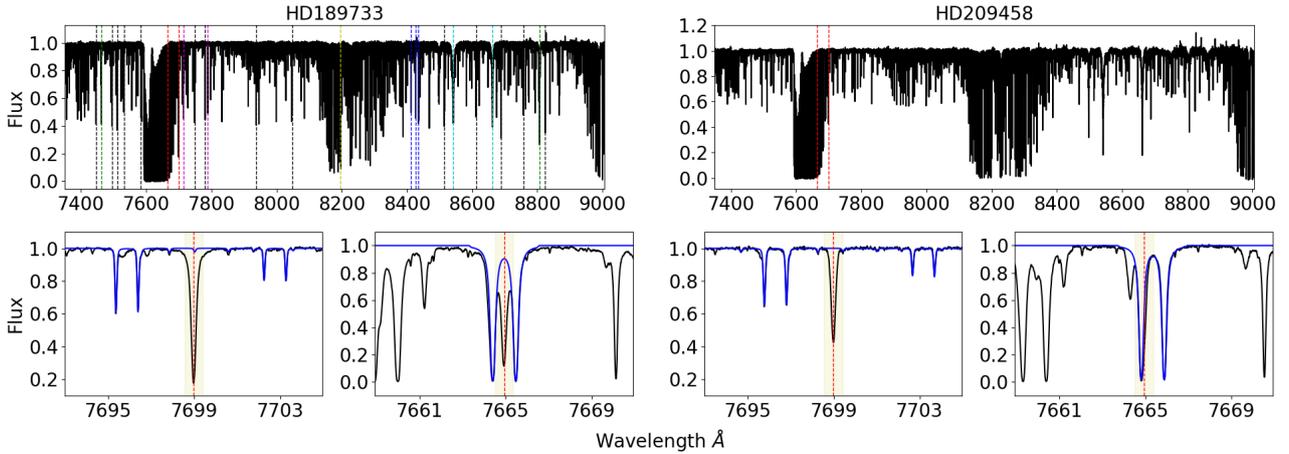

**Figure 1.** Full PEPSI spectra (top) and K lines (bottom). Dashed lines mark the K lines (red) and control lines (colored) and the solid blue line mark the telluric model for few telluric lines. Shaded area shows the planetary motion of around ±0.42Å during the transit.

(PI: K.G. Strassmeier, AIP) using the PEPSI instrument (Strassmeier et al. 2015, 2018b) with a 3.2– pixel resolution of 130 000 at the LBT operated in the binocular mode. The wavelength region of interest was covered with cross disperser VI (7340 Å – 9070 Å) in the polarimetric mode and the beams recombined yielding the integral light spectrum. The spectrograph is a white–pupil fiber–fed spectrograph located in a pressure-controlled chamber at a constant pressure, temperature and humidity to ensure constant refractive index of the air inside, providing radial velocity stability about 5 m/s on the long term and less than 0.5 m/s on the short term (Strassmeier et al. 2018a).

For the HD189733b transit, we obtained 24 spectra (15 out-of-transit (OOT) and 9 in-transit (IT)) setting the exposure time to 10 min. We exclude one OOT exposure due to a systematically lower relative flux determination. The resulting signal-to-noise ratio (S/N) in the final processed data varied due to increase in airmass during the night in the K line at 7699 Å from around 160 to 70 and at the continuum at 7700 Å from around 540 to 270. For the HD209458b transit, we obtained 17 spectra (9 OOT and 8 IT) setting the exposure time to 10 min. The observations paused from 06:26 UT – 07:19 UT due to bad weather conditions, leading to a loss of phase coverage at the second part of the transit. The resulting S/N in the final processed data varied due to an increase in airmass during the night in the K line at 7699 Å from around 230 to 80 and at the continuum at 7700 Å from around 425 to 160.

The image processing steps includes bias subtraction and variance estimation of the source images, super–master flat field correction for the CCD spatial noise, echelle orders definition from the tracing flats, scattered light subtraction, wavelength solution for the ThAr images, optimal extraction of image slicers and cosmic spikes elimination of the target image, wavelength calibration and merging slices in each order, normalization to the master flat field spectrum to remove CCD fringes and blaze function, a global 2D fit to the continuum of the normalized image, and rectification of all spectral orders in the image to a 1D spectrum. The software numerical toolkit and graphical interface application for PEPSI is described in Strassmeier et al. (2018a), and its complete description is in preparation by Ilyin (2019).

The blaze function was removed by the division of the master flat field spectrum. The residual spectra were fitted with a low order 2D spline with subsequent rectification of spectral orders where the overlapping parts of spectral orders were averaged with their weights as the inverse variances of the individual wavelength pixels. After that, the continuum of individual spectra was corrected again with the mean spectrum as the weighted average of all the observed spectra. The mean spectrum was normalized to its continuum and each observed spectrum was divided by the mean normalized spectrum. A low order spline fit to the ratio constitutes the final continuum of the individual spectrum.

## 3 DATA ANALYSIS

The K doublet absorption lines are at 7698.98 Å and 7664.92 Å (see bottom panels of Figure 1). Since the latter line is surrounded by a strong pair of telluric oxygen lines, we focus our analysis only on the line at 7698.98 Å. To demonstrate the reliability of the excess absorption, we apply the same procedure on several control lines, where we do not expect planetary absorption (see e.g. Redfield et al. (2008)). Each chosen control line is free of telluric contamination and has a normalized flux level < 0.5 in the line centre. We use synthetic stellar spectra to model the center-to-limb variation (CLV) (see Section 3.3) to avoid a false-positive detection. Top panels in Figure 1 show PEPSI spectra of the targets.

### 3.1 Telluric lines

Telluric lines are spectral features induced by Earths atmosphere. To investigate if telluric line contamination has an effect on the results (as the HD189733 spectra exhibit a weak telluric line at 7699 Å), each spectrum is telluric corrected around the K lines (see bottom panel of Figure 1) and then sampled to a common wavelength grid using a spline function. For this, we developed a telluric line corrector called "Telluric Hapi Observation Reducer" (THOR), which uses the "HITRAN Application Programming Interface" (HAPI) as a basis (Kochanov et al. 2016). HAPI consists of different python routines enabling the calculation of absorption spectra using line-by-line data provided by the HITRAN database (Gordon et al. 2017). THOR iterates HAPI for different parameters (e.g. wavelength shifts, wavelength resolution, mean free path) until it reaches a $\chi^2$ – min between the observed and modelled telluric lines. To prove the reliability of our telluric correction, we verify the increase in line depth with airmass for the modelled telluric line at 7699 Å applying a Pearson correlation test, which result in a value of 0.90 (thus showing high correlation). As the telluric contamination is weak, the excess level shown in Section 4 do not change within 1-$\sigma$ either applying telluric correction or not. For further analysis, we use the telluric corrected spectra.

### 3.2 Excess absorption

For both targets, the planetary motion during the transit of around ±16 km/s introduces a shift in the absorption wavelength of approximately ±0.42Å. To search for the K excess absorption, we use bandwidths from 0.8 Å to 10.0 Å in steps of 0.4 Å (whereby one pixel corresponds to ∼0.016 mÅ). This lower limit of the bandwidth ensures that a major part of the planetary absorption (shifted by the planetary motion) is inside the integration band. The lower limit also contains a major part of the spectral line, avoiding artifacts introduced by line shape changes (Snellen et al. 2008) due to the Rossiter- McLaughlin- effect.

We use two methods to infer the planetary excess absorption. First, we follow the "traditional" way, integrating the flux in the spectral range of interest and dividing it by the mean of the integrated flux (at same bandwidth) at two reference positions adjacent to the red and blue. The center of the reference bands for the HD189733b investigation are at 7693.40 Å and 7700.40 Å and for the HD209458b investigation at 7692.40 Å and 7705.40 Å. Second, we will infer the excess absorption by using no reference bands at all (thus only integrating the flux in the K-line). This is possible for high S/N continuum normalized spectra with stable blaze functions. For both of these methods, we made use of a 2-order polynomial fit to the OOT data for normalization.

The error bars are calculated using the uncertainties sourcing from the photon and readout noise propagated in the data. They are scaled according to the standard deviation of the OOT values, if the mean error was underestimated compared to this standard deviation. We determine the excess absorption level using a Markov-Chain Monte Carlo (MCMC) method provided by PyAstronomy (a collection of Python routines implemented in the PyMC (Patil et al. 2010) and SciPy packages (Jones et al. 01 ), where we simulate a transit and scale the area of a planetary body to mimic an absorbing atmosphere. We applied 100 000 MCMC iterations (rejecting the first 30000 samples as burn-in phase) to ensure that the final best-fit is provided. We checked for convergence of the chains by splitting them in three equally sized sub-groups and verified that their individual mean agreed within 1-$\sigma$, whereby their individual uncertainty agree within 3.5%. The uncertainties for the excess absorption level correspond to a 1-$\sigma$ confidence level.

### 3.3 Center-to-limb variation

As the planet covers different parts of the stellar surface during its transit, the differential limb darkening between the line core and the stellar continuum leads to darkening or brightening effects, which are evident in the excess absorption curves. A detailed discussion of the CLV effect and its influence on excess absorption curves is shown e.g. in Czesla et al. (2015) and Yan et al. (2017). We simulate a transit to derive the CLV curves using synthetic stellar spectra, which are calculated using the "spectrum" program by R.O. Gray (Gray & Corbally 1994). For the calculation of the model atmospheres we used the Kurucz model (Kurucz 1970; Castelli & Kurucz 2004). For HD189733b, we used an effective temperature of 4875 K, a surface gravity of log g = 4.56 and a metallicity of dex -0.03. For HD209458b, we used an effective temperature of 6092 K, a surface gravity of log g = 4.25 and a metallicity of dex 0.02 (see also Boyajian et al. (2015) for comparison). We set $\mu = \cos\theta$ to specify the limb angle, whereby $\mu = 0$ refers to the limb and $\mu = 1$ to the center of the disk. We generated 20 spectra with limb angles between $\mu = 0$ and $\mu = 1$ with a spacing of 0.05. To derive the CLV curves, the stellar surface is mapped by a grid of $100 \times 100$ pixels containing the limb angle dependent fluxes and the planetary surface is mapped by $31.6 \times 31.6$ pixels (whereby higher pixel values did not change significantly the results, but increased the computational time). For each planet position in front of the star, we calculate the total stellar limb angle dependent flux at the position where we expect planetary absorption and the reference band position. The CLV- curve is then produced by the same way as the excess curve. We validated our model by comparing the CLV effect for the sodium excess absorption simulated for HD189733b by Yan et al. (2017), getting similar results.

## 4 RESULTS

### 4.1 Investigating potassium on HD189733b

The left panel of Figure 2 shows the mean excess absorption curve for HD189733b at a 0.8 Å integration band using adjacent reference bands (top) and no reference bands (middle). The blue curve shows the MCMC fit and the green solid line represents the CLV effect. As the CLV effect has a negligible effect on the overall excess absorption, we neglect it in the middle panel. The excess absorption levels are 0.181 % ± 0.022 % (top) and 0.184 % ± 0.025 % (middle), thus similar within their error bars. This shows that both methods are suitable to infer the planetary excess absorption. This absorption value corresponds to ∼13 scale heights, hinting that the absorption must originate at high altitudes in the atmosphere. The left bottom panel shows the mean absorption for different bandwidths (using no reference bands) and the significance level. By increasing the bandwidth more than 0.8 Å, the absorption level decreases, as expected, due to the integration of less atmospheric K absorption relative to the continuum flux. Also shown on the right y-scale is the apparent planetary radius i.e. the radius until which the atmosphere is opaque in units of the white light radius. The significance level of the K detection for the 0.8 Å bandwidth is determined with $> 7$-$\sigma$ with respect to the zero level. For the remaining other bandwidths, the determined excess absorption levels are also above 3-$\sigma$ with respect to the zero level.

Comparing our results to other investigations, we can qualitatively confirm the tentative K- detection of Pont et al.

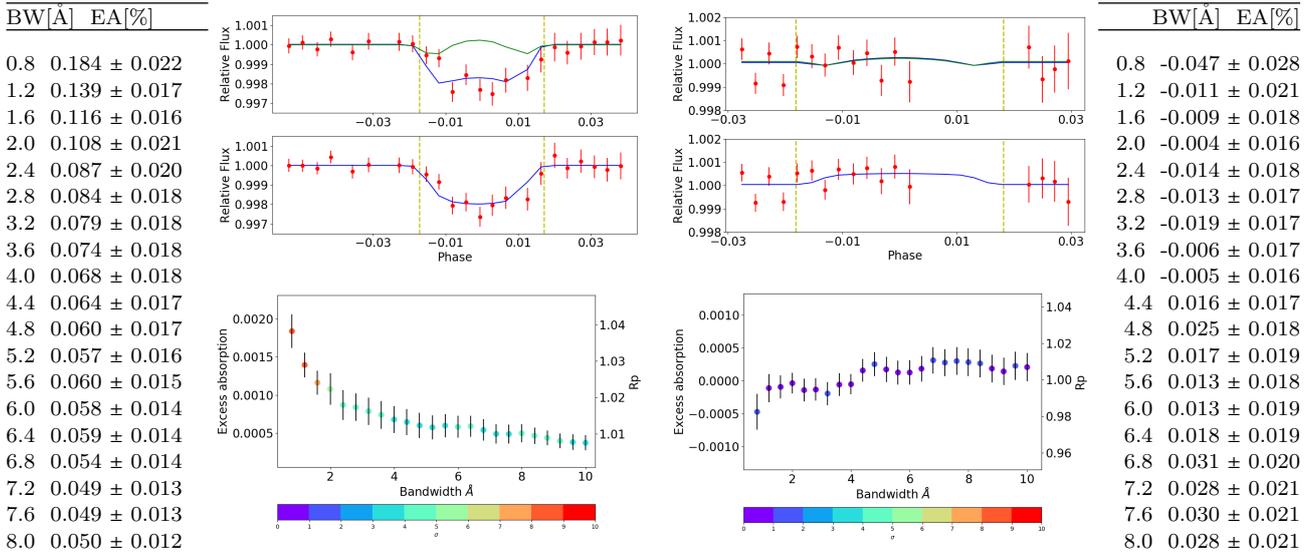

**Figure 2.** K excess absorption (EA) for HD189733b (left) and HD209458b (right). EA curves at 0.8 Å bandwidth (BW) using reference bands (top) and no reference bands (middle). Bottom panel shows the EA level for different BW. Green line shows the CLV-curve and the blue line the planetary absorption (both mirrored to mimic the second half of the transit for HD209458b). Tables show EA up to 8Å.

(2013). Other way around, Jensen et al. (2011) investigated at a resolution of 60 000 potassium on HD189733b, stating a non-detection. In their work, they averaged the IT and OOT spectra from several observations to produce one master $F_{in}/F_{out}$ spectrum and determined thereafter from this the excess absorption. To do this, they subtracted the integrated flux from the line core at the K-line from the mean of two reference band fluxes using an 8Å integration band deriving an excess level of $(-0.77 \pm 1.04) \times 10^{-4}$. Determining the excess absorption at 8Å bandwidth, we infer an excess absorption value of $(4.95 \pm 1.20) \times 10^{-4}$, deviating significantly more than 3-$\sigma$ from their findings. Possible reasons which could explain this discrepancy are e.g. telluric lines and stellar activity in some of the observations which can affect the combined master spectra or even the difference in the technique used to derive the excess level.

Figure 3 shows the excess level at a 0.8 Å integration band (using no reference bands) for several control lines which all lie within 3-$\sigma$ around zero opposite to the excess level at the K- line at 7698.98 Å (red square). The significance level of the K absorption remains > 3-$\sigma$ with respect to the standard deviation of the excess level for the control lines, presenting a strong evidence on the atmospheric K- absorption.

### 4.2 Stellar activity of HD189733

As the star HD189733 is an active K-dwarf, the stellar variability could cause errors in the determined excess levels e.g. by flares (Klocová et al. 2017; Khalafinejad et al. 2017; Cauley et al. 2018). Cauley et al. (2018) showed for HD189733b that the transit of active latitudes with bright facular and plage regions can cause emission feature in stellar lines. Also the strong magnetic field of HD189733 could have a significant effect on the emission in lines (Cauley et al. 2017). Possibly related to the activity, we see an emission-like feature in several stellar line cores as e.g. in Na I (8183.28 Å), Fe I (7511.03 Å) and the investigated K I line (7698.98 Å). As the integrated flux in the line core is low compared to the integrated flux over the line width (of at least 0.8Å), this has a negligible influence on the result. Moreover, this feature increases over the night and it is not restricted to the transit, opposite to the excess absorption in Figure 2, which appears only during the transit, making us confident about the K absorption within the atmosphere of HD189733b.

### 4.3 Investigating potassium on HD209458b

The right panel in Figure 2 shows the same approach on HD209458b as for HD189733b. Similar to HD189733b, the CLV- effect is very weak and does not affect the result. In contrast, there is no excess absorption evident at the investigated bandwidths, either using reference bands or not. Moreover, the result suggests an emission like behaviour at low bandwidths. Assuming zero excess absorption to not underestimate the error (as a negative excess level would suggest emission either than absorption), we determine a 3-$\sigma$ upper limit of around 0.084% at a bandwidth of 0.8Å. Concluding, in comparison to HD189733b, HD209458b shows no significant absorption of potassium in its atmosphere.

## 5 DISCUSSION

Although both targets could have experienced different evolutionary scenarios, changing their primordial atmospheres leading to different atmospheric composition and properties, we make the assumption that the initial alkali metal abundances could be similar for both targets, as they orbit a host star of solar metallicity (Boyajian et al. 2015). As gas giants form and accrete H/He-dominated gas, they also accrete planetesimals that enrich their envelope in metals, thus they can possess even a higher absolute metellicity compared to the parent star (Nikolov et al. 2018). Independent of any formation location in the protoplanetary disc, the alkali metal ratios should not be affected significantly, as the accretion of planetesimals will enrich both (here Na and K) similarly. Therefore, to make an assumption about the expected K excess level on HD209458b, we compare the previously detected Na excess level on both planets assuming similar Na/K– ratio. Although the Na excess absorption level for HD189733b is around two times larger than for HD2093458 (Snellen et al. 2008; Jensen et al. 2011), it corresponds to a similar variation in apparent planet radius (Snellen et al. 2008). Translating the K excess absorption of 0.184% on HD189733b to the variation in apparent planet radius, one would expect a K absorption of around 0.108% for HD209458b. This value is larger than our upper 3-$\sigma$ limit of 0.084%, indicating that K could be depleted on HD209458b, either by condensation processes in the lower and/or photo-ionization processes in the upper atmosphere. Such an indication is also suggested by Sing et al. (2008b,a), who observed a broad absorption plateau at lower altitudes and a narrow absorption line core for the Na feature. The authors argued that condensation on the night side of the planet can lead to the loss of atmospheric Na, as the atmospheric temperatures become cool enough to condensate Na and lead to the absence on high altitudes. This was also confirmed by Vidal-Madjar et al. (2011) showing temperature-pressure-profile simulations to match the observations by Sing et al. (2008b,a) and Snellen et al. (2008).

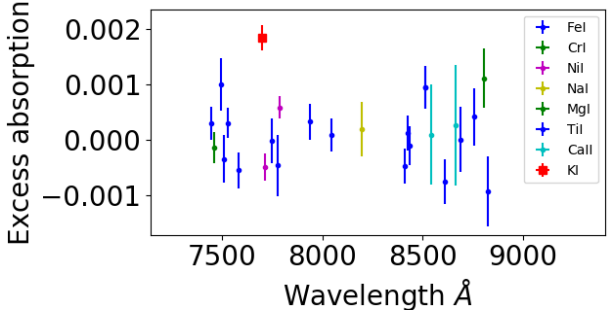

**Figure 3.** Mean excess absorption for several control lines compared to K (red square) at 0.8 Å bandwidth for HD189733b.

Comparing to those findings, the interpretation of the K detection is puzzling. The condensation of Na to $Na_2S$ as well as NaCl happens at higher temperature than the condensation of K to KCl (Lodders 1999), so that one would expect K to be more abundant than Na, in contrast with what is observed. An alternative depletion process is photo-ionization. Potassium has a slightly lower ionization energy than Na, leading to easier photo-ionization of K (Fortney et al. 2003; Barman 2007). Then, one may expect that depletion of K is stronger on HD189733b (which orbits an active star) than on HD209458b, not being in agreement with our findings, making the atmospheric conditions on HD209458b puzzling.

Although both planets have similar properties (e.g. size, orbital period and equilibrium temperature) there are deviations especially on their cloud properties (Lines et al. 2018). Modelling approaches suggest that cloud formation on HD189733b originates at lower pressure levels due to its higher gravity and cloud particle density (Lines et al. 2018) opposite to dust clouds on HD209458b, which extent to large areas on the atmosphere (Helling et al. 2016).

These differences (among others) could lead to different condensation chemistry on both targets, affecting the alkali depletion processes and thus the implications discussed above.

This paper has been typeset from a TeX/LaTeX file prepared by the author.